\documentclass[
twocolumn,aps,prb,
 amsmath,amssymb,
floatfix,superscriptaddress
]{revtex4-2}

\usepackage{graphicx}
\usepackage{epstopdf}
\usepackage{physics}
\usepackage{bbold}
\usepackage{amsmath}
\usepackage[colorlinks = true,
            linkcolor = blue,
            urlcolor  = blue,
            citecolor = blue,
            anchorcolor = blue]{hyperref}

\begin{document}
\title{Subharmonic spin correlations and spectral pairing in Floquet time crystals}
\author{Alexander-Georg Penner}
\affiliation{\mbox{Dahlem Center for Complex Quantum Systems, Freie Universit\"at Berlin, 14195 Berlin, Germany}}
\author{Harald Schmid}
\affiliation{\mbox{Dahlem Center for Complex Quantum Systems, Freie Universit\"at Berlin, 14195 Berlin, Germany}}
\author{Leonid I.\ Glazman}
\affiliation{\mbox{Department of Physics and Yale Quantum Institute, Yale University, New Haven, Connecticut 06520, USA}}
\author{Felix von Oppen}
\affiliation{\mbox{Dahlem Center for Complex Quantum Systems, Freie Universit\"at Berlin, 14195 Berlin, Germany}}
\begin{abstract}
Floquet time crystals are characterized by subharmonic behavior of temporal correlation functions. Studying the paradigmatic time crystal based on the disordered Floquet quantum Ising model, we show that its temporal spin correlations are directly related to spectral characteristics and that this relation provides analytical expressions for the correlation function of finite chains, which compare favorably with numerical simulations. Specifically, we show that the disorder-averaged temporal spin correlations are proportional to the Fourier transform of the splitting distribution of the pairs of eigenvalues of the Floquet operator, which differ by $\pi$ to exponential accuracy in the chain length. We find that the splittings are  well described by a log-normal distribution, implying that the temporal spin correlations are characterized by two parameters. We discuss possible implications for the phase diagram of the Floquet time crystals. 

\end{abstract}
\maketitle
\date{\today}
\section{Introduction}
There has recently been much interest in the possibility of spontaneous breaking of time translation symmetry \cite{Wilczek2012,Sacha2018,Khemani2019,Choi2017,Zhang2017,Randall2021,Mi2022,Frey2022,Else2020,Zaletel2023}. While spontaneous breaking of spatial translation symmetries underlies equilibrium crystalline phases, time translation symmetry breaking has only been established under out-of-equilibrium conditions. This has led to the concept of Floquet time crystals with the Floquet quantum Ising chain providing a paradigmatic model  \cite{Khemani2016,Else2016}. Randomness in the exchange couplings as well as in integrability-breaking fields introduces localization, which suppresses thermalization and enables subharmonic temporal spin correlations.  

The subharmonic correlations are closely related to spectral properties of the Floquet operator. In the case of the Floquet quantum Ising model, its eigenstates appear in pairs with eigenphases that differ by $\pi$ (taking the period of the Floquet drive equal to $T=1$). While this pairing becomes exact in the thermodynamic limit, there are deviations from $\pi$ for finite chains. As discussed in Refs.\ \cite{Keyserlingk2016a,Surace2019}, the typical deviations are exponentially small in the system size. Here, we show that the probability distribution of these splittings away from $\pi$ is accurately described by a log-normal distribution. Moreover, the disorder-averaged  temporal spin correlation function of the time crystal is directly proportional to the Fourier transform of the splitting
distribution, providing an analytical description of the temporal spin correlations and their evolution with chain length. 

Earlier studies of spectral diagnostics of Floquet time crystals focused on statistics of adjacent energy levels and measures of entanglement \cite{Lazarides2015,Khemani2016,Yao2017,Lazarides2015,Ponte2015,Bairey2017,Sonner2021,Sierant2023}. While thermalizing systems tend to follow Wigner-Dyson-like level repulsion, the absence of level repulsion is closely related with obstructions to thermalization. Unlike these more generic measures of spectral correlations, the distribution of $\pi$ pairings is directly related to the defining temporal spin correlations of discrete time crystals. The deviations $\Delta$ from perfect $\pi$ pairing lead to a decay of the subharmonic spin correlations at times, which are exponentially long in the system size. Our results show that the distribution of $\pi$ pairings directly controls how the time-crystalline state is established in the thermodynamic limit. Moreover, the log-normal distribution of $\Delta$ entails that temporal spin correlations of finite-size systems are characterized by two parameters, namely the average and the variance of  $\ln\Delta$. 

Our results are based on a combination of numerical simulations and analytical arguments for a disordered Floquet quantum Ising chain, both without and with longitudinal random field. The appearance of a log-normal distribution can be traced to the fact that the deviations from $\pi$ pairings are related to tunneling processes traversing the entire Floquet quantum Ising chain.
The two partner states are even and odd eigenstates of the spin-flip symmetry of the Ising model. In the limit of weak transverse fields, their splitting reflects tunneling processes between two oppositely polarized eigenstates of the ferromagnet. Effectively, this occurs by propagation of a domain wall around the system and the associated amplitude is a product of a large number of random factors. 

Alternatively, this can be viewed from the fermionized version of the model \cite{Lieb1961,Pfeuty1979}. In this formulation, the two partner states have opposite fermion parity. The sectors with even and odd fermion parity (known as Neveu-Schwarz and Ramond sectors) satisfy antiperiodic and periodic boundary conditions of the Jordan-Wigner fermions, respectively. Consequently, the splittings reflect the sensitivity to boundary conditions. In localized systems, the sensitivity to boundary conditions is frequently characterized by log-normal distributions, as for the conductance \cite{Imry1986}, the level curvatures \cite{Titov1997}, or the splitting of end states \cite{Brouwer2011}.

 \begin{figure*}[t!]
\includegraphics[width=\textwidth]{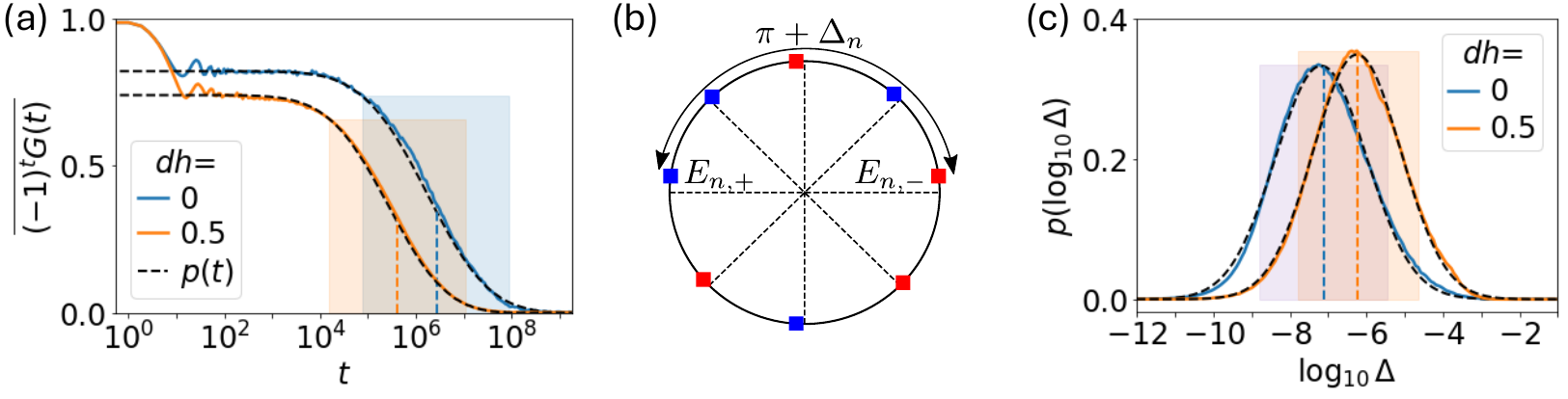}
\caption{(a) Disorder average of the bulk spin correlation function $G(t)$ (solid lines) of a discrete time crystal without (blue) and with (orange) longitudinal fields. Dashed lines: Fourier transform $p(t)$ of a log-normal distribution $p(\Delta)$ with cumulants extracted from the splitting distribution. $p(t)$ has been multiplied by an overall prefactor to match the plateau value of $G(t)$. Shaded boxes: decay range estimated from $p(\Delta)$. (b) Illustration of many-body Floquet spectrum with $\pi$-paired partner states. Even and odd states under the spin-flip symmetry are indicated by blue and red. Sketch also indicates the finite-size splittings $\Delta_n$ away from $\pi$. (c) Numerical splitting distributions (solid lines) without (blue) and with (orange) longitudinal fields, compared to log-normal fits (dashed lines). The splitting distribution is sampled across many-body spectrum and disorder ensemble. Shaded boxes: characteristic range determining time scales of spin correlations in (a), see text for further discussion. Parameters: $N=12$, $g=0.95$, $dJ=0.25$, $J=0.5$, $\mathcal{N}=10^{3}$ disorder realizations.}
\label{fig: correlation function}
\end{figure*}

The remainder of this paper is structured as follows. In Sec.\ \ref{sec:numerical}, we use numerical simulations to establish a relation between the temporal spin correlations of the Floquet time crystal with the splitting distribution of the $\pi$ pairs. 
Section \ref{sec:analytical} supports our numerical results by analytical considerations. In particular,
we analyze the splitting distribution within the original spin model in Sec.\ \ref{sec:spinmodel} and within the framework of Jordan-Wigner fermions in Sec.\ \ref{sec:fermionmodel}. The effects of an integrability-breaking longitudinal field are discussed in the framework of a self-consistent perturbation theory in Sec.\ \ref{sec:randlongfield}. We conclude in Sec.\ \ref{sec:conclusions}.

\section{Temporal spin correlations and spectral pairing: Numerical results}
\label{sec:numerical}

We study the Floquet quantum Ising chain with random exchange couplings and random longitudinal field as a model for discrete time crystals. The Floquet operator  
\begin{align}
\label{eq:U_F}
U_F = e^{\frac{i\pi g }{2}\sum\limits^N_{j=1} X_j}
   e^{\frac{i\pi }{2}\sum\limits^{N}_{j=1}J_j Z_j Z_{j+1}} 
     e^{\frac{i\pi  }{2}\sum\limits^N_{j=1} h_jZ_j }
\end{align}
defines the stroboscopic time evolution $\ket{\Psi(t)}=U^t_{F} \ket{\Psi(0)}$ of an initial state $\ket{\Psi(0)}$ 
after  $t=0,1,2,\dots$ cycles. 
Here, $X_j$ and $Z_j$ are Pauli operators at site $j$, and we assume periodic boundary conditions, i.e., $Z_{N+1} = Z_{1}$. We draw the Ising couplings $J_j$ and longitudinal fields $h_j$ from independent box distributions, $J_j \in [J-dJ,J+dJ]$ and $h_j \in [-dh,dh]$. The Floquet operator can be diagonalized, $U_{F}\ket{n}=e^{-iE_n}\ket{n}$, with  eigenphases $E_n$ and Floquet eigenstates $\ket{n}$. The eigenphases can be restricted to the first Floquet zone, $-\pi \leq E_n \leq \pi$. The quantum Ising model ($dh=0$) has a spin-flip symmetry $P=\prod_j X_j$ and is integrable. The random longitudinal field breaks integrability. Spin-flip symmetry is broken for particular realizations of the longitudinal field, but maintained on average by the disorder ensemble. 

Time-crystalline behavior occurs for transverse fields $g$ close to one \cite{Khemani2016,Else2016}. We characterize the dynamics of the time crystal by the infinite-temperature correlation function  
\begin{equation}
 \label{eq: cor fun}
     G_j(t) = \expval{ Z_j(t)Z_j(0)} ,
\end{equation}
where $Z_j(t)=(U_F^\dagger)^tZ_j(U_F)^t $ and $\expval{\dots}=2^{-N}\mathrm{tr}[\dots]$ averages over a complete set of states. We focus on the disorder-averaged correlation function $\overline{(-1)^tG(t)}$ (as denoted by the overline). The disorder average makes the correlation function translationally invariant for periodic boundary conditions, so that we drop the site index $j$. At $g=1$ and $dh=0$, the transverse field flips all spins periodically, resulting in perfect period doubling as diagnosed by $G(t)=(-1)^t$. Away from $g=1$, the transverse field no longer induces complete spin flips. Provided that $g$ does not deviate too far from $g=1$, one finds that following initial transients, 
$(-1)^tG(t)$ plateaus for many cycles before slowly decaying to zero at very long times. This characteristic behavior remains qualitatively unchanged in the presence of the random longitudinal field. 
Figure \ref{fig: correlation function}(a) shows corresponding numerical results both without and with random longitudinal field. 

The period doubling characteristic of discrete time crystals originates from a spectrum-wide pairing of eigenstates  [Fig.\ \ref{fig: correlation function}(b)]. For $dh=0$, the partner states $\ket{n,\pm}$  are even $(+)$ and odd $(-)$ under the spin-flip symmetry $P$ and have eigenphases $E_{n,\pm}$ differing by $\pi$ up to finite-size corrections, which are exponentially small in the length $N$ of the chain. The operators $Z_j$ are odd under the spin-flip symmetry and couple partner states, resulting in the period doubling of $G(t)$. This follows from writing Eq.\ \eqref{eq: cor fun} in terms of exact eigenstates,
\begin{equation}
    G_j(t) = \frac{1}{2^N}\sum_{nm} |\langle n |Z_j| m\rangle |^2 e^{i(E_n-E_m)t}.
\end{equation} 
This equation also implies that provided they are inhomogeneous across the spectrum, the finite-size deviations from $\pi$ pairing cause the time-crystal oscillations to decay at exponentially long times. 

To probe the relation between spectral pairing and time-crystal oscillations more closely, we compute the deviations $\Delta_n=\pi-E_{n,+}+E_{n,-}$ of the splittings from $\pi$ across the Floquet spectrum and the disorder ensemble by exact diagonalization. Since paired eigenphases are almost antipodal, they can be identified by sorting the spectrum provided that the splittings are small compared to the level spacing. We show below that this condition is satisfied for transverse fields, which are sufficiently close to $g=1$, see Eq.\ \eqref{eq:goodsplit}. We characterize the splittings by their distribution function across the many-body spectrum and the disorder ensemble, which is shown in Fig.\ \ref{fig: correlation function}(c), both without and with longitudinal field. We find that in both cases, the splitting distribution is well fit by a log-normal distribution 
\begin{align}
    p\left(\ln  \Delta \right) \dd\ln \Delta =     \frac{1}{\sqrt{2\pi}\sigma}e^{-\frac{1}{2\sigma^2}(\ln \Delta-\mu)^2 } \dd\ln \Delta.
\end{align}
Small deviations from the log-normal distribution appear in the tails of the distribution and are more pronounced without random longitudinal field. 

\begin{figure}
    \centering
    \includegraphics[width=0.8\linewidth]{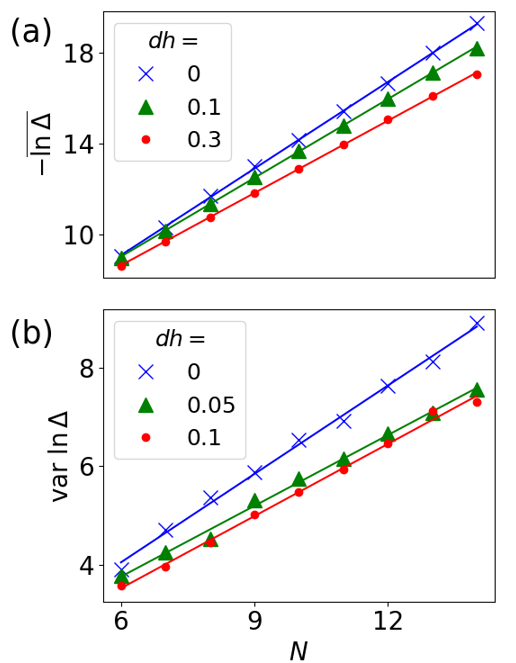}
    \caption{Dependence on chain length $N$ of (a) mean and (b) variance of many-particle splittings, both without and with random longitudinal field (see legends). Symbols: data;  solid lines: linear fits. Parameters: $J=0.5$, $dJ=0.25$, $g=0.95$. Disorder averages over $\mathcal{N}=5\cdot 10^3$ realizations ($dh>0$, $N=6,\dots,13$), $\mathcal{N}=2\cdot 10^3$  ($dh>0$, $N=14$), $\mathcal{N}=2\cdot 10^4$ ($dh=0$, all $N$). } 
    \label{fig: mean var}
\end{figure}

The log-normal distribution is fully determined by the first two cumulants of $\ln \Delta$. Figure \ref{fig: mean var} shows the scaling of these cumulants with system size $N$. We read off that
\begin{align}
\mu = \overline{\ln \Delta }=-\frac{N}{\zeta}, \qquad \sigma^2=\mathrm{var} \ln \Delta=\frac{N}{\lambda}.
\label{eq:musigscalingN}
\end{align}
The average $\overline{\ln \Delta }$ (characterizing
typical values of $\Delta$) decreases linearly in the chain length $N$. In contrast, the variance $\mathrm{var} \ln \Delta$ grows with $N$. The prefactors of $N$ define characteristic lengths $\zeta$ and $\lambda$, respectively. Figure \ref{fig: correlation function}(c) shows that as a function of the random longitudinal field, the average increases (shift to the right), while the variance  shrinks (reduced width and increased peak height).

It is our central result that beyond initial transients,
the Fourier transform 
\begin{equation}
p(t) = \int \dd{\Delta} p(\Delta) e^{i \Delta t}
\label{eq:ptFTpd}
\end{equation}
of the log-normal splitting distribution closely tracks the correlation function $\overline{(-1)^tG(t)}$. Figure \ref{fig: correlation function}(a) shows that  $p(t)$ (dashed lines) matches $\overline{(-1)^tG(t)}$ (full lines) up to an overall prefactor. This can be understood based on Eq.\ \eqref{eq: cor fun}. The period-two oscillations arise from the contributions of the $\pi$-paired levels. If fluctuations of the positive matrix-element prefactor across the spectrum or the disorder ensemble do not affect the temporal spin correlations,  
Eq.~(\ref{eq: cor fun}) yields
\begin{equation}
\overline{(-1)^tG(t)} \propto  \frac{1}{2^N}  \overline{\sum_{n} e^{i(\pi - E_{n,+} + E_{n,-}) t} }.
\label{eq:ptFTpd2}
\end{equation}
This relates the temporal spin correlations to the Fourier transform of the splitting distribution. 

The relation between 
Eq.\ \eqref{eq:ptFTpd} and the temporal spin correlations implies that the latter are controlled by the two cumulants $\mu$ and $\sigma$ of the log-normal distribution. In particular, we can express the two characteristic times of the spin correlation function in terms of these cumulants. 
The log-normal distribution takes on appreciable values for $\exp{\mu- \sqrt{2}\sigma} \lesssim \Delta \lesssim \exp{\mu +\sqrt{2} \sigma}$. Correspondingly, its Fourier transform remains constant for $t \lesssim t^* = \exp{ -\mu - \sqrt{2} \sigma}$ and falls off over the interval between $t^*$ and  $ t^{**} =  \exp{ -\mu + \sqrt{2} \sigma} $, as illustrated by the shaded boxes in Figs.\ \ref{fig: correlation function}(a) and (c). In combination with Eq.\ \eqref{eq:musigscalingN} this implies that time-crystalline behavior, i.e., a nonzero $\overline{G(t)}$, persists for all times, when taking the thermodynamic limit before the $t \to \infty$ limit. We can also check that the ratio between ``maximal" splitting $\exp{\mu + \sqrt{2}\sigma}$ and many-body level spacing $\propto 2^{-N}$ vanishes for $N\to\infty$ as long as
\begin{equation}
    \zeta < 1/\ln 2,
\label{eq:goodsplit}
\end{equation} i.e., as long as $g$ is sufficiently close to unity as mentioned above, ensuring that the splitting distribution is well defined.  

An approximate analytical expression for $p(t)$ can be obtained by evaluating the Fourier transform over the log-normal distribution within the saddle-point approximation. Keeping only quadratic deviations from the saddle, one finds \cite{Asmussen2016}
\begin{align}
\label{eq:saddle Fourier}
    p(t)&\simeq 
    \Re
    \frac{\exp\left\{-\frac{1}{2\sigma^2}\left[W^2(-i\sigma^2 e^\mu t)+2W(-i\sigma^2e^\mu t)\right] \right\} }{\sqrt{1+W(-i\sigma^2 e^\mu t)}}
    ,
\end{align}
where $W$ is the Lambert-W function defined by  $W(x)e^{W(x)}=x$. Equation \eqref{eq:saddle Fourier} is controlled in the limit of small $\sigma$, but reproduces the Fourier transform qualitatively even when $\sigma$ is of order unity. One also deduces the approximate expression $p(t) \simeq e^{-(1/(2\sigma^2))\ln^2(\sigma^2 e^\mu t)}$ for the asymptote at large $t$. 

\section{Analytical considerations}
\label{sec:analytical}

\subsection{Spin model}
\label{sec:spinmodel}

We explore the spectral properties underlying these numerical results by analytical considerations in limiting cases. First focus on the Floquet operator $U_{F,0} = U_F(h_j=0)$
in the absence of the random longitudinal field $h_j$, 
\begin{align}
\label{eq:U_F0}
U_{F,0} = P e^{\frac{i\pi \delta g }{2}\sum\limits^N_{j=1} X_j}
   e^{\frac{i\pi }{2}\sum\limits^{N}_{j=1}J_j Z_j Z_{j+1}}.
\end{align}
Since we are interested in Floquet operators with transverse field $g$ close to one, we have pulled out an overall factor of $P = \prod_j X_j$ by writing $g=1 + \delta g$ with $\delta g\ll 1$. The Floquet evolution commutes with the spin-flip symmetry, so that we can consider the even ($P=1$) and odd ($P=-1$) sectors of the model separately. In the limit of $\delta g \to 0$, the eigenstates of $U_{F,0}$ for $P=\pm 1$ are even and odd linear combinations of oppositely polarized bit strings, e.g., 
\begin{equation}
    \frac{1}{\sqrt{2}}(\ket{\uparrow \downarrow \downarrow \ldots \downarrow} \pm \ket{\downarrow \uparrow \uparrow \ldots \uparrow}).
    \label{eq:evenoddstatesschem}
\end{equation}
Due to the overall factor of $P$, the corresponding eigenvalues are $\pm e^{iE}$ with antipodal eigenphases $E$ and $E+\pi$. 

Splittings away from $\pi$ appear for nonzero $\delta g$. We consider the regime of intermediate disorder strength,
\begin{equation}
    \delta g \ll dJ \ll J \ll 1.
    \label{eq:limit}
\end{equation}
This allows us to approximate the Floquet operator as
\begin{align}
\label{eq:U_F0approx}
U_{F,0} \simeq  P e^{\frac{i\pi \delta g }{2}\sum\limits^N_{j=1} X_j + \frac{i\pi }{2}\sum\limits^{N}_{j=1}J_j Z_j Z_{j+1}},
\end{align}
where the exponent is the Hamiltonian of the transverse field Ising model with a small transverse field. Thus, the splittings can be obtained from ordinary Hamiltonian perturbation theory in $\delta g$. 

The splittings arise from tunneling between the two oppositely polarized bit strings in Eq.\ \eqref{eq:evenoddstatesschem} induced by spin flips of amplitude $\delta g$. Since $J$ is largest, all exchange couplings are positive and, to leading order, the number of domain walls remains unchanged in the virtual intermediate states. Considering a state with two domain walls  (the smallest number compatible with periodic boundary conditions) for definiteness, a tunneling event can be thought of as a process, in which the two domain walls trade locations in such a way that the $N$ hops effectively make a full loop around the system. The corresponding splitting appears in $N$th order perturbation theory with amplitude  
\begin{equation}
    \Delta_{I,K} \simeq \frac{\pi \delta g}{2} \sum_\gamma  \prod_{(i,j)\in \gamma} \frac{\delta g}{J_I + J_K - J_i - J_j},
    \label{eq:DeltaIK}
\end{equation}
where $I,K$ denote the locations of the domain walls of the $\pi$ pair. The sum is over all contributing trajectories $\gamma$, with $(i,j)$ running through the $N-1$ intermediate  configurations with domain walls located at $i$ and $j$.

A log-normal splitting distribution is a plausible consequence of the appearance of products with an extensive number of factors in Eq.\ \eqref{eq:DeltaIK}. Provided that the sum over $\gamma$ is dominated by sufficiently few terms and that the factors can be viewed as statistically independent, $\ln\Delta$ becomes normally distributed by the central limit theorem. The small deviations of the splitting distribution for $dh=0$ from a log-normal distribution, as shown in Fig.\ \ref{fig: correlation function}(c), indicate the degree to which these assumptions are justified. 

The two degenerate ground states of the transverse field Ising model in the limit of zero transverse field contain no domain walls. Thus, their splitting is due to tunneling processes with intermediate states of excitation energy $\sim J$ and consequently, in the limit under consideration, much smaller than the splitting of excited states. 

We remark that the assumption of $J\ll 1$ was made to simplify the presentation. None of our results depend on it in essential ways. In fact, we can use a stroboscopic Floquet perturbation theory \cite{schmid2024robust} (see also Sec.\ \ref{sec:randlongfield} below) to discuss the eigenphases for general $J$, with the same qualitative results. 

\begin{figure*}[t!]
\includegraphics[width=\textwidth]{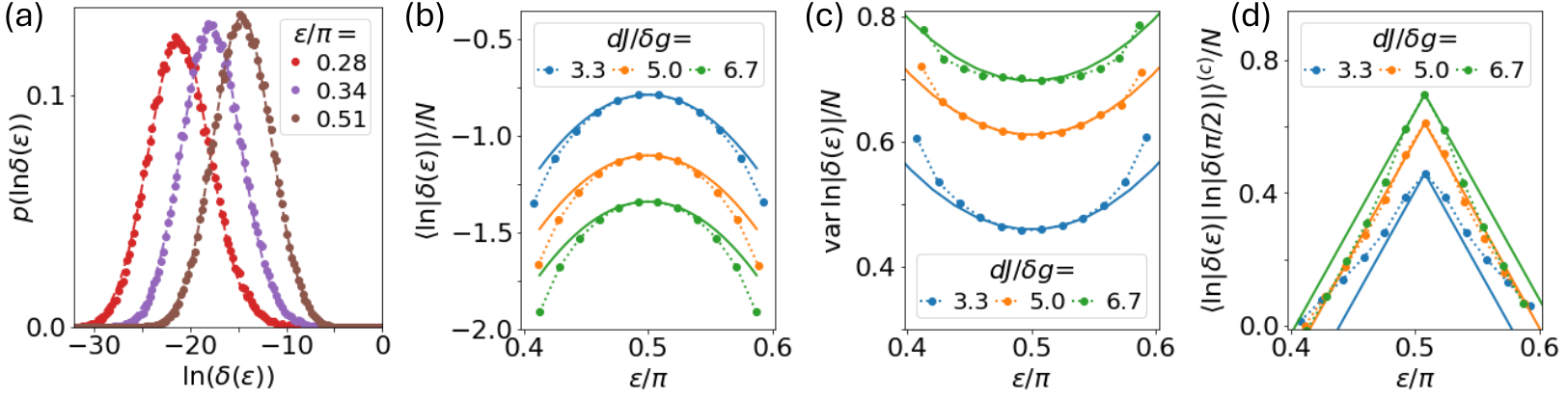}
\caption{Single-particle splittings $\delta(\epsilon)$ for different boundary conditions of the Jordan-Wigner fermions and various energies $\epsilon$ in the band. (a) Distributions across the disorder ensemble (dots). Dashed lines: log-normal fits. (b) Average, (c) variance and (d) correlations of $\ln(\delta(\epsilon))$ (dotted). Solid lines: Perturbation theory. Lines have been shifted by a constant.     
Parameters: $J=0.5$, $\mathcal{N} = 10^5$ disorder realizations, (a) $dJ = 0.25$, $N=18$, $g=0.9$,  (b)-(d) $dJ = 0.1$, $N=12$ (for $g$ see legends).}
\label{fig: single-particle splittings}
\end{figure*}

\subsection{Fermionized model}
\label{sec:fermionmodel}

For zero random longitudinal field ($dh=0$), the Floquet Ising model is integrable and its many-body spectrum is composed of single-particle levels. We can investigate the distribution of $\pi$ pairings from the viewpoint of these single-particle levels.

A Jordan-Wigner transformation maps the Floquet quantum Ising model to a free-fermion model in two ways (see App.\ \ref{app:JWT}) \cite{Lieb1961,Pfeuty1979}. Applying the Jordan-Wigner transformation to the original spin operators, the fermion operators $c_j$ effectively describe spin flips in the transverse-field basis and the Floquet operator takes the form
\begin{align}
    U_{F,0} = P
   &\exp\left\{\frac{i\pi \delta g}{2}\sum\limits_{j=1}^N (1-2c^\dagger_jc_j)\right\}
    \nonumber
   \\ 
   \times & \exp\left\{ -\frac{i\pi}{2}\sum\limits_{j=1}^N J_j (c^\dagger_{j}c_{j+1}+c^\dagger_{j}c^\dagger_{j+1}) 
   +\mathrm{h.c.}\right\}.
\label{fermionfloquetoperator}
\end{align}
Here, we identify $c_{N+1} = - P c_1$. Thus, while the original spin model has periodic boundary conditions, the fermionized model has periodic (antiperiodic) boundary conditions for $P=-1$ ($P=1$). This difference between the Neveu-Schwarz ($P=-1$) and Ramond ($P=1$) sectors of the model appears since the Jordan-Wigner string in $Z_NZ_1$ winds around the entire chain. In the fermionic representation, the spin-flip symmetry maps $P$ to the fermion parity
$P= \prod_{j}e^{i\pi c^\dagger_jc_j}$.  

In the limit of small $\delta g$, it is advantageous to apply the Jordan-Wigner transformation to the dual spin operators describing domain walls. The corresponding fermion operators 
\begin{equation}
    d_j^\dagger = \frac{1}{2}(c_j + c_j^\dagger + c_{j+1} - c_{j+1}^\dagger)
\end{equation}
effectively describe domain walls. In terms of these operators, the Floquet operator takes the form
\begin{align}
    U_{F,0} = P
   &\exp\left\{\frac{i\pi \delta g}{2}\sum\limits_{j=1}^N (d_j + d_j^\dagger) (d_{j-1} -  d_{j-1}^\dagger)\right\}
    \nonumber
   \\ 
   \times & \exp\left\{ \frac{i\pi}{2}\sum\limits_{j=1}^N J_j (1 - 2 d_j^\dagger d_j) 
   \right\}.
\end{align}
Analogous to the convention for the spin-flip fermions $c_j$, the domain-wall fermions satisfy $d_0 = - P d_N$ (see App.\ \ref{app:JWT}). Due to the noninteracting nature of this fermion problem, the eigenphases of $U_{F,0}$ can be written as $E[\{n_\alpha\}] = \sum_\alpha n_\alpha \epsilon_\alpha$, where $n_j$ are fermionic occupation numbers and $\epsilon_\alpha$ are single-particle eigenphases.

In the limit defined in Eq.\ \eqref{eq:limit}, we can combine the exponentials, so that apart from the factor $P$, $U_{F,0}$ reduces to the time-evolution operator of \begin{equation}
   H = \frac{\pi\delta g }{2} \sum\limits_{j=1}^N (d_j + d_j^\dagger) (d_{j-1}^\dagger -  d_{j-1}) + \pi  \sum\limits_{j=1}^N J_j  d_j^\dagger d_j. 
\end{equation}
The contributions of the pairing terms (corresponding to the generation or annihilation of pairs of domain walls) are parametrically suppressed in a perturbative expansion in $\delta g$, so that the Hamiltonian reduces to a noninteracting Anderson model with random on-site disorder. 
Moreover, we have already seen in Sec.\ \ref{sec:spinmodel} that to leading order, we can neglect the dependence of the ground-state energy on $P$. Thus, the splittings of the $\pi$ pairs arise from the dependence of the single-particle excitations on $P$.  

The Anderson models for $P =  \pm 1$ effectively differ by a $\pi$ flux threading the ring, or equivalently, in their boundary conditions. As argued by Thouless \cite{Thouless1977}, the sensitivity of the single-particle levels to the boundary conditions is a measure of the conductance of the system. Moreover, the conductance of a one-dimensional disordered wire obeys a log-normal distribution \cite{Imry1986} as a consequence of Oseledec's theorem \cite{Oseledec1968}. We thus conclude that the single-particle splittings $\delta_\alpha = \epsilon_\alpha(P=+1)-\epsilon_\alpha(P=-1)$
have a log-normal distribution. This expectation is in excellent agreement with the numerical data in Fig.\ \ref{fig: single-particle splittings}(a).

These conclusions can be understood as follows. As a consequence of Anderson localization, the corresponding difference in the single-particle energies is exponentially small in $N/\xi(\epsilon)$, where $\xi(\epsilon)$ is the localization length at the energy $\epsilon$ of the single-particle excitation. In zeroth order in the transverse field, the single-particle spectrum consists of fermions localized at sites $j$ with energy $\pi J_j$, so that $\epsilon/\pi \in [J-dJ,J+dJ]$. A nonzero transverse field introduces hopping between nearest-neighbor sites. The splitting between the $P=\pm 1$ sectors is due to hopping around the entire chain, which  appears in $N$th-order perturbation theory. Thus, the splitting of the excitation with $\epsilon = \pi J_j$ is given by the amplitude
\begin{equation}
    \delta_j \simeq 2 \pi \delta g \prod_{l(\neq j)} \frac{\pi \delta g/2}{\epsilon - \pi J_l}. \label{fermionfloquetoperatordual}
\end{equation}
This expression is a limiting case of an exact relation with a similar structure \cite{Thouless1972}. 

Unlike in the perturbative approach to the many-body eigenphases $E_n$ in Sec.\ \ref{sec:spinmodel}, only two terms of equal amplitude contribute to Eq.\ \eqref{fermionfloquetoperatordual} at this order of perturbation theory and the factors are statistically independent. The log-normal distribution is thus a direct consequence of the central limit theorem. As for the many-body eigenphases, we can characterize the splitting distribution of the single-particle eigenphases by 
\begin{align}
    \overline{\ln \delta(\epsilon)}=-\frac{N}{\xi(\epsilon)}, \quad \mathrm{var}\ln \delta(\epsilon)=\frac{N}{\ell(\epsilon)},
\end{align}
in terms of the localization length $\xi(\epsilon)$ and the elastic mean free path $\ell(\epsilon)$ of the Anderson model. Using Eq.\ \eqref{fermionfloquetoperatordual} for $N\gg 1$ and small $\epsilon -\pi J$, we find  \cite{Thouless1979}
\begin{equation}
    \overline{ \ln |\delta(\epsilon)|}
    \simeq -N\left[\ln\frac{2 dJ}{e\delta g} +\frac{(\epsilon-\pi J)^2}{2(\pi dJ)^2}\right]
\end{equation}
and 
\begin{equation}
     \mathrm{var} \ln |\delta(\epsilon)|
     \simeq N \left[1+\frac{(\epsilon-\pi J)^2}{(\pi dJ)^2}\right].
\end{equation}
The expression for $ \overline{ \ln |\delta(\epsilon)|}$ shows that  states near the band edges are most strongly localized. Correspondingly, their splitting fluctuations $\mathrm{var} \ln |\delta(\epsilon)|$ are  largest. The dependence of these quantities on $\epsilon $ is in good agreement with numerical results within the range of validity of the expansion in $\epsilon-\pi J$, see Fig.\ \ref{fig: single-particle splittings}(b) and (c). We remark that the limit defined in Eq.\ \eqref{eq:limit} is challenging to realize numerically. Deviations necessitate an additional overall vertical shift of the fitting curves in Figs.\ \ref{fig: single-particle splittings}(b) and (c). Consistent with expectations, the shift diminishes when approaching the limit in Eq.\ \eqref{eq:limit}.

The splitting $\Delta$ of many-body eigenphases is given as a sum of many log-normally distributed single-particle splittings $\delta$. Since a log-normal distribution has a finite average and variance, one may be tempted to apply the central-limit theorem and conclude that $\Delta$ should have a normal distribution, in contradiction to the observed log-normal distribution. This conundrum is resolved by noting that the single-particle splittings violate the assumption of statistical independence underlying the central limit theorem. In fact, the single-particle splittings are correlated even for states with widely differing energies. We can use Eq.\ \eqref{fermionfloquetoperatordual} to compute the (connected) 
correlations of the single-particle splittings at different energies. Working to first order in $\epsilon-\epsilon^\prime$, we find 
\begin{align}
     \overline{\langle  \ln |\delta(\epsilon)| \ln |\delta(\epsilon^\prime)| \rangle}^{(c)}
     \simeq N\left[1-\frac{|\epsilon-\epsilon^\prime|\ln^2(\pi dJ)}{2\pi dJ} \right].
\end{align}
This result is in good agreement with numerical results as shown in Fig.\ \ref{fig: single-particle splittings}(d) and explains the kink at $\epsilon = \epsilon^\prime$. The correlations drop on a scale $\epsilon_c=\frac{2\pi dJ}{\ln^2(\pi dJ)}$, so that the single-particle splittings are correlated over a wide range of energies. This precludes the application of the central limit theorem to the sum over single-particle splittings and explains how the log-normal distribution of the many-body splittings emerges from the log-normally distributed single-particle splittings.

\begin{figure}[t!]
\raggedright
\includegraphics[scale=0.5]{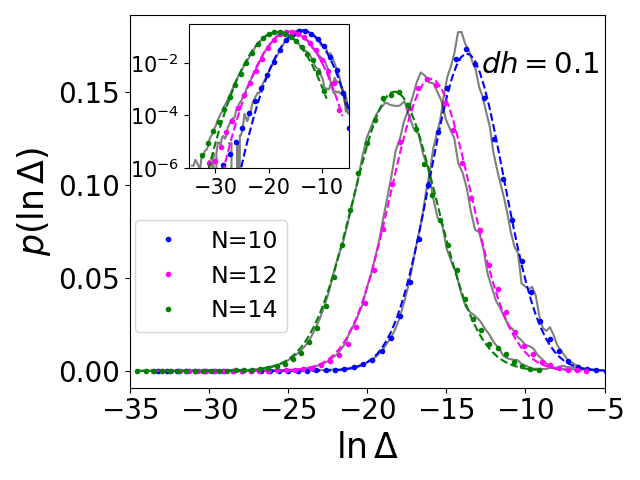}
\caption{Distribution of the many-body splittings with disordered longitudinal field for various chain lengths (see legend). Numerical data (symbols) are compared to the self-consistent perturbation theory (solid gray lines). Dashed line: Log-normal fits to data. Inset: Same on a doubly-logarithmic scale. Parameters: $J=0.5$, $dJ=0.25$, $g=0.95$, $\mathcal{N}=10^4$ disorder realizations (for $N=10,12$) and $\mathcal{N}=10^4$ ($N=14$).}
\label{fig: splittings longitudinal field}
\end{figure}

\subsection{Random longitudinal field}
\label{sec:randlongfield}

We now turn to a discussion of the random longitudinal field, which breaks integrability as well as the global spin-flip symmetry $P$. As shown in Fig.\ \ref{fig: correlation function}(a), the log-normal splitting distribution persists in the presence of a random longitudinal field. We find that the mean grows with the strength of the longitudinal field, while the width in $\ln \Delta$ shrinks (with a corresponding increase in peak height). As shown in Fig.\ \ref{fig: mean var}, the mean and the width of the distribution in $\ln\Delta$ continue to depend exponentially on chain length $N$, with the length scales $\xi(dh)$ and $\ell(dh)$ in Eq.\ \eqref{eq:musigscalingN} increasing with field.

The effect of the random longitudinal field can be understood within a stroboscopic Floquet
perturbation theory for $U_F=U_{F,0}e^{iV}$, where we treat  $V=\sum_jh_jZ_j$ as a perturbation \cite{schmid2024robust}. Expanding the eigenphases $E_n = E_{n,0}+E_{n,1}+E_{n,2}+\dots$ of $U_F$ to quadratic order in $V$ gives
\begin{equation}
     E_{n,1} = \langle n_0| V |n_0 \rangle
\end{equation}
and 
\begin{equation}     E_{n,2} =\sum_{m \neq n} \frac{|\matrixel{n_0}{V}{m_0}|^2}{2\tan \frac{E_{n,0}-E_{m,0}}{2}}. 
     \label{eq:pt}
\end{equation}
Here, the unperturbed eigenstates $\ket{n_0}$ of $U_{F,0}$ are 
assumed to be nondegenerate with eigenphases $E_{n,0}$. We denote the unperturbed $\pi$ pairs as $\ket{n_0,\pm}$ with $P|n_0, \pm \rangle =\pm |n_0, \pm \rangle$ and unperturbed many-body splittings by
\begin{align}
    \Delta_{n_0}=\pi-E^+_{n_0}+E^-_{n_0}.
\end{align}
The longitudinal field is odd under $P$, so that it couples states with opposite parity. As a result, the first-order contribution vanishes and Eq.\ \eqref{eq:pt} gives the perturbed splittings 
\begin{align}
\label{eq:splitting pt}
    \Delta_n\simeq \Delta_{n_0}-\sum_{m}\left[ \frac{|v^{+-}_{nm}|^2}{2\tan \frac{E_{n,+}-E_{m,-}}{2}}
    -
    \frac{|v^{-+}_{nm}|^2}{2\tan \frac{E_{n,-}-E_{m,+}}{2}}\right]
\end{align}
with matrix elements  $v^{+-}_{nm}=\mel{n,+}{V}{m,-}$.

In Eq.\ \eqref{eq:splitting pt}, we have made the perturbation theory self consistent by inserting the exact eigenphases into the denominators \cite{schmid2024robust}. This is motivated as follows. While the divergent eigenphase denominator suppresses the contribution of $\pi$ pairs (i.e., the terms with $m=n$), there are many terms of similar magnitude with $m\neq n$. This is plausibly captured by a self-consistent approximation. An analogous scheme was  applied successfully to the related problem of Majorana $\pi$ modes in chains with open boundary conditions \cite{schmid2024robust}. 

We make the splittings $\Delta_n = \pi-E^+_{n}+E^-_{n}$ explicit in the energy denominators and expand in them as they are small compared to the eigenphase differences. To linear order, we find
\begin{equation}
    \Delta_n \simeq  \Delta_{n,0} +\Lambda_n -\Sigma_n \Delta_n
\end{equation}
in terms of
\begin{align}
        \Lambda_n &\simeq  \sum_m\left[\frac{|v_{nm}^{-+}|^2}{2\tan\frac{E^+_{n,0}-E^+_{m,0}+\pi}{2}}-\frac{|v_{nm}^{+-}|^2}{2\tan\frac{E^-_{n,0}-E^-_{m,0}+\pi}{2}}\right].
        \label{eq: Lambda}
        \\
    \Sigma_n &\simeq   \sum_m\left[\frac{|v_{nm}^{+-}|^2}{4\cos^2\frac{E^-_{n,0}-E^-_{m,0}}{2}}+\frac{|v_{nm}^{-+}|^2}{4\cos^2\frac{E^+_{n,0}-E^+_{m,0}}{2}}\right], 
    \label{eq: Sigma}
\end{align}
In defining $\Sigma_,\Lambda_n$, we assume that the exact eigenphases $E^\pm_{n}$ can now be replaced by their unperturbed counterparts $ E^\pm_{n,0}$. Solving for $\Delta_n$, we find
\begin{equation}
    \Delta_n \simeq \frac{\Delta_{n,0} + \Lambda_n}{1+\Sigma_n}.
    \label{eq:self-cons sol}
\end{equation}
similar to an analogous expression in Ref.\ \cite{schmid2024robust}. 

Figure \ref{fig: splittings longitudinal field} shows that the splitting distributions in the presence of random longitudinal field are well reproduced by the self-consistent perturbation theory. Deviations from the log-normal fits appear deep in the tails of the distribution, as shown in the inset. 

\section{Conclusion}
\label{sec:conclusions}

Floquet time crystals with subharmonic response at half the driving frequency are known to exhibit $\pi$ pairing in their eigenphase spectrum \cite{Keyserlingk2016a,Surace2019}. In finite-size systems, the eigenphase difference splits away from $\pi$ by an exponentially small amount. We showed that the probability distribution of these splittings across the many-body spectrum and the disorder ensemble is well approximated by a log-normal distribution. Moreover, its Fourier transform is directly proportional to the temporal spin correlations of finite-size time crystals. This provides an immediate relation between the spectral statistics of the Floquet operator and the defining characteristic of Floquet time crystals. 

Our results open various avenues for further research. First, it is an interesting question whether other spin correlation functions can be related to the spectral statistics of $\pi$ pairings. In particular, recent work has emphasized the role of out-of-time-order correlators of Floquet time crystals \cite{Sahu2024}. Second, it is important to understand whether a similar relation between spectral statistics and time-crystalline order extends to Floquet time crystals with other subharmonic periods \cite{Surace2019,Pizzi2019,Pizzi2021,Giachetti2023} to time crystalline behavior in open quantum systems \cite{Iemini2018,Tucker2018,Zhu2019,Gambetta2019,Booker2020,Krishna2023}, or disorder-free models \cite{Van2019,Kshetrimayum2020,Liu2023}. Third, our work relies on a combination of numerical simulations and analytical arguments. It would be interesting to see whether the relation between spin correlations and splitting distributions as well as the latter's log-normal nature can be made rigorous.

Our results may also have implications for the phase diagram of Floquet time crystals. It is natural to speculate that the relation between the pairing distribution and time-crystalline spin correlations is a basic property of the time-crystal phase. This would suggest that the time-crystal phase requires the spectral $\pi$ pairing to be well defined, i.e., that the splittings are small compared to the many-body level spacing of order $2^{-N}$. Under these assumptions, the condition in Eq.\ \eqref{eq:goodsplit} for the existence of well-defined splittings describes the phase boundary of the time-crystal phase. It would thus be interesting to understand the dependence of the characteristic length $\zeta$ on system parameters. This line of thinking may also have ramifications for the existence of time crystals in higher than one dimension. 

\begin{acknowledgments}
We thank R.\ Fazio, S.\ Sondhi, and K.\ Yang for helpful discussions. We gratefully acknowledge funding by Deutsche Forschungsgemeinschaft through CRC 183 as well as the Einstein Research Unit on Quantum Devices (FvO), and by NSF Grant No.\ DMR-2410182 (LG). We thank the HPC service of ZEDAT, Freie
Universität Berlin, for computing time \cite{Bennett2020}.
\end{acknowledgments}

\appendix

\section{Jordan-Wigner transformation}
\label{app:JWT}

We use the Jordan-Wigner transformation
\begin{equation}
    X_j = 1 - 2c_j^\dagger c_j^{\phantom{\dagger}}  \quad ; \quad Z_j = i( c_j^\dagger - c_j^{\phantom{\dagger}} ) \,\exp\left\{i\pi \sum_{l<j} c_l^\dagger c_l^{\phantom{\dagger}} \right\}  .
\end{equation}
This implies
\begin{equation}
    Y_j = -iZ_jX_j =  (c_j + c_j^\dagger) \exp\left\{i\pi \sum_{l<j} c_l^\dagger c_l^{\phantom{\dagger}} \right\},
\end{equation}
so that the spin raising and lowering operators   $S_j^\pm = \frac{1}{2}(Y_j \pm i Z_j)$ become
\begin{equation}
    S_j^+  = c_j e^{i\pi \sum_{l<j} c_l^\dagger c_l^{\phantom{\dagger}} } \quad ; \quad S_j^- = c_j^\dagger e^{i\pi \sum_{l<j} c_l^\dagger c_l^{\phantom{\dagger}} }. 
\end{equation}
We associate $c_j^\dagger$ with the spin-lowering operator $S_j^-$ as the latter creates a spin-flip excitation in the paramagnetic phase. 

We introduce Majorana operators defined through
\begin{equation}
    c_j = \frac{1}{2}(a_j + i b_j), 
\end{equation}
so that 
\begin{equation}
    X_j  = - i a_j b_j 
    \, ; \,
    Y_j  = a_j \prod_{l<j} (-ia_l b_l) \, ; \,
    Z_j  =  b_j \prod_{l<j} (-ia_l b_l) .
    \nonumber
\end{equation}
We can express the Floquet quantum Ising model with periodic boundary conditions in terms of the Majorana operators using
\begin{equation}
    g \sum_{j=1}^N X_j = - i g  \sum_{j=1}^N  a_j b_j
\end{equation}
and 
\begin{equation}
    \sum_{j=1}^N J_j Z_j Z_{j+1} = i \sum_{j=1}^N J_j a_j b_{j+1}  .
\end{equation}
We note that $Z_NZ_1 = - i P a_Nb_1 $, which we accounted for by the conventions $c_{N+1} = - P c_1$ as well as $a_{N+1} = - P a_1$ and $b_{N+1} = - P b_1$. Note that in these conventions, $P=\pm 1$ denotes the sector of Hilbert space and is no longer an operator. 

In terms of the spin-flip operators $c_j$, we have 
\begin{equation}
    g \sum_{j=1}^N X_j = g  \sum_{j=1}^N  (1-2c_j^\dagger c_j)
\end{equation}
and 
\begin{equation}
    \sum_{j=1}^N J_j Z_j Z_{j+1} =  \sum_{j=1}^N J_j (c_j+c_j^\dagger) (c_{j+1}-c_{j+1}^\dagger).
\end{equation}
In this formulation, spin flips cost a Zeeman energy of $2g$, while the exchange coupling either hops the spin flip to a neighboring site, or creates or annihilates a pair of spin flips on nearest-neighbor sites.

We can also define fermions for which the exchange term becomes diagonal, 
\begin{equation}
    d_j^\dagger = \frac{1}{2}(a_j + i b_{j+1}). 
\end{equation}
These are associated with the dual representation of the quantum Ising model in terms of domain-wall operators. In fact, $d_j^\dagger d_j$ counts the number of domain walls at the bond between sites $j$ and $j+1$, 
\begin{equation}
    \sum_{j=1}^N J_j Z_j Z_{j+1} =  \sum_{j=1}^N J_j (1 - 2 d_j^\dagger d_j).
    \label{eq:gXdop1}
\end{equation}
Moreover, the transverse-field term turns into
\begin{equation}
    g \sum_{j=1}^N X_j = g  \sum_{j=1}^N  (d_j + d_j^\dagger) (d_{j-1} -  d_{j-1}^\dagger).
    \label{eq:gXdop2}
\end{equation}
In this formulation, domain walls cost an exchange energy of $2J$, while the Zeeman coupling either hops the domain walls to a neighboring bond, or creates or annihilates a pair of domain walls on nearest-neighbor bonds.

We finally discuss the influence of the fermion parity $P$ on the domain-wall representation. Recalling the sign convention for the fermion operators, we note that 
\begin{equation}
    d_N^\dagger = \frac{1}{2} ( a_N + i b_{N+1}) = \frac{1}{2} ( a_N - i b_1 P),
\end{equation}
so that 
\begin{equation}
    i J_N a_N b_{N+1} = - i J_N P a_N b_1  = J_N ( 1 - 2  d_N^\dagger d_N)
\end{equation}
and 
\begin{equation}
    -i g a_1 b_1 = gP (d_1 + d_1^\dagger)(d_N^\dagger - d_N).
\end{equation}
This is consistent with Eqs.\ \eqref{eq:gXdop1} and \eqref{eq:gXdop2} if we use the convention $d_0 = - P d_N$.


\begin{thebibliography}{45}%
\makeatletter
\providecommand \@ifxundefined [1]{%
 \@ifx{#1\undefined}
}%
\providecommand \@ifnum [1]{%
 \ifnum #1\expandafter \@firstoftwo
 \else \expandafter \@secondoftwo
 \fi
}%
\providecommand \@ifx [1]{%
 \ifx #1\expandafter \@firstoftwo
 \else \expandafter \@secondoftwo
 \fi
}%
\providecommand \natexlab [1]{#1}%
\providecommand \enquote  [1]{``#1''}%
\providecommand \bibnamefont  [1]{#1}%
\providecommand \bibfnamefont [1]{#1}%
\providecommand \citenamefont [1]{#1}%
\providecommand \href@noop [0]{\@secondoftwo}%
\providecommand \href [0]{\begingroup \@sanitize@url \@href}%
\providecommand \@href[1]{\@@startlink{#1}\@@href}%
\providecommand \@@href[1]{\endgroup#1\@@endlink}%
\providecommand \@sanitize@url [0]{\catcode `\\12\catcode `\$12\catcode
  `\&12\catcode `\#12\catcode `\^12\catcode `\_12\catcode `\%12\relax}%
\providecommand \@@startlink[1]{}%
\providecommand \@@endlink[0]{}%
\providecommand \url  [0]{\begingroup\@sanitize@url \@url }%
\providecommand \@url [1]{\endgroup\@href {#1}{\urlprefix }}%
\providecommand \urlprefix  [0]{URL }%
\providecommand \Eprint [0]{\href }%
\providecommand \doibase [0]{http://dx.doi.org/}%
\providecommand \selectlanguage [0]{\@gobble}%
\providecommand \bibinfo  [0]{\@secondoftwo}%
\providecommand \bibfield  [0]{\@secondoftwo}%
\providecommand \translation [1]{[#1]}%
\providecommand \BibitemOpen [0]{}%
\providecommand \bibitemStop [0]{}%
\providecommand \bibitemNoStop [0]{.\EOS\space}%
\providecommand \EOS [0]{\spacefactor3000\relax}%
\providecommand \BibitemShut  [1]{\csname bibitem#1\endcsname}%
\let\auto@bib@innerbib\@empty
\bibitem [{\citenamefont {Wilczek}(2012)}]{Wilczek2012}%
  \BibitemOpen
  \bibfield  {author} {\bibinfo {author} {\bibfnamefont {F.}~\bibnamefont
  {Wilczek}},\ }\href {\doibase 10.1103/PhysRevLett.109.160401} {\bibfield
  {journal} {\bibinfo  {journal} {Phys. Rev. Lett.}\ }\textbf {\bibinfo
  {volume} {109}},\ \bibinfo {pages} {160401} (\bibinfo {year}
  {2012})}\BibitemShut {NoStop}%
\bibitem [{\citenamefont {Sacha}\ and\ \citenamefont
  {Zakrzewski}(2017)}]{Sacha2018}%
  \BibitemOpen
  \bibfield  {author} {\bibinfo {author} {\bibfnamefont {K.}~\bibnamefont
  {Sacha}}\ and\ \bibinfo {author} {\bibfnamefont {J.}~\bibnamefont
  {Zakrzewski}},\ }\href {\doibase 10.1088/1361-6633/aa8b38} {\bibfield
  {journal} {\bibinfo  {journal} {Rep. Prog. Phys.}\ }\textbf {\bibinfo
  {volume} {81}},\ \bibinfo {pages} {016401} (\bibinfo {year}
  {2017})}\BibitemShut {NoStop}%
\bibitem [{\citenamefont {Khemani}\ \emph {et~al.}(2019)\citenamefont
  {Khemani}, \citenamefont {Moessner},\ and\ \citenamefont
  {Sondhi}}]{Khemani2019}%
  \BibitemOpen
  \bibfield  {author} {\bibinfo {author} {\bibfnamefont {V.}~\bibnamefont
  {Khemani}}, \bibinfo {author} {\bibfnamefont {R.}~\bibnamefont {Moessner}}, \
  and\ \bibinfo {author} {\bibfnamefont {S.~L.}\ \bibnamefont {Sondhi}},\
  }\href@noop {} {\enquote {\bibinfo {title} {A brief history of time
  crystals},}\ } (\bibinfo {year} {2019}),\ \Eprint
  {http://arxiv.org/abs/1910.10745} {arXiv:1910.10745} \BibitemShut {NoStop}%
\bibitem [{\citenamefont {{S. Choi \textit{et al.}}}(2017)}]{Choi2017}%
  \BibitemOpen
  \bibfield  {author} {\bibinfo {author} {\bibnamefont {{S. Choi \textit{et
  al.}}}},\ }\href {https://doi.org/10.1038/nature21426} {\bibfield  {journal}
  {\bibinfo  {journal} {Nature}\ }\textbf {\bibinfo {volume} {543}},\ \bibinfo
  {pages} {221} (\bibinfo {year} {2017})}\BibitemShut {NoStop}%
\bibitem [{\citenamefont {{J. Zhang \textit{et al.}}}(2017)}]{Zhang2017}%
  \BibitemOpen
  \bibfield  {author} {\bibinfo {author} {\bibnamefont {{J. Zhang \textit{et
  al.}}}},\ }\href {https://doi.org/10.1038/nature21413} {\bibfield  {journal}
  {\bibinfo  {journal} {Nature}\ }\textbf {\bibinfo {volume} {543}},\ \bibinfo
  {pages} {217} (\bibinfo {year} {2017})}\BibitemShut {NoStop}%
\bibitem [{\citenamefont {Randall}\ \emph {et~al.}(2021)\citenamefont
  {Randall}, \citenamefont {Bradley}, \citenamefont {van~der Gronden},
  \citenamefont {Galicia}, \citenamefont {Abobeih}, \citenamefont {Markham},
  \citenamefont {Twitchen}, \citenamefont {Machado}, \citenamefont {Yao},\ and\
  \citenamefont {Taminiau}}]{Randall2021}%
  \BibitemOpen
  \bibfield  {author} {\bibinfo {author} {\bibfnamefont {J.}~\bibnamefont
  {Randall}}, \bibinfo {author} {\bibfnamefont {C.~E.}\ \bibnamefont
  {Bradley}}, \bibinfo {author} {\bibfnamefont {F.~V.}\ \bibnamefont {van~der
  Gronden}}, \bibinfo {author} {\bibfnamefont {A.}~\bibnamefont {Galicia}},
  \bibinfo {author} {\bibfnamefont {M.~H.}\ \bibnamefont {Abobeih}}, \bibinfo
  {author} {\bibfnamefont {M.}~\bibnamefont {Markham}}, \bibinfo {author}
  {\bibfnamefont {D.~J.}\ \bibnamefont {Twitchen}}, \bibinfo {author}
  {\bibfnamefont {F.}~\bibnamefont {Machado}}, \bibinfo {author} {\bibfnamefont
  {N.~Y.}\ \bibnamefont {Yao}}, \ and\ \bibinfo {author} {\bibfnamefont
  {T.~H.}\ \bibnamefont {Taminiau}},\ }\href {\doibase 10.1126/science.abk0603}
  {\bibfield  {journal} {\bibinfo  {journal} {Science}\ }\textbf {\bibinfo
  {volume} {374}},\ \bibinfo {pages} {1474} (\bibinfo {year}
  {2021})}\BibitemShut {NoStop}%
\bibitem [{\citenamefont {{X. Mi \textit{et al.}}}(2022)}]{Mi2022}%
  \BibitemOpen
  \bibfield  {author} {\bibinfo {author} {\bibnamefont {{X. Mi \textit{et
  al.}}}},\ }\href {https://doi.org/10.1038/s41586-021-04257-w} {\bibfield
  {journal} {\bibinfo  {journal} {Nature}\ }\textbf {\bibinfo {volume} {601}},\
  \bibinfo {pages} {531} (\bibinfo {year} {2022})}\BibitemShut {NoStop}%
\bibitem [{\citenamefont {Frey}\ and\ \citenamefont {Rachel}(2022)}]{Frey2022}%
  \BibitemOpen
  \bibfield  {author} {\bibinfo {author} {\bibfnamefont {P.}~\bibnamefont
  {Frey}}\ and\ \bibinfo {author} {\bibfnamefont {S.}~\bibnamefont {Rachel}},\
  }\href {\doibase 10.1126/sciadv.abm7652} {\bibfield  {journal} {\bibinfo
  {journal} {Science Advances}\ }\textbf {\bibinfo {volume} {8}},\ \bibinfo
  {pages} {eabm7652} (\bibinfo {year} {2022})}\BibitemShut {NoStop}%
\bibitem [{\citenamefont {Else}\ \emph {et~al.}(2020)\citenamefont {Else},
  \citenamefont {Monroe}, \citenamefont {Nayak},\ and\ \citenamefont
  {Yao}}]{Else2020}%
  \BibitemOpen
  \bibfield  {author} {\bibinfo {author} {\bibfnamefont {D.~V.}\ \bibnamefont
  {Else}}, \bibinfo {author} {\bibfnamefont {C.}~\bibnamefont {Monroe}},
  \bibinfo {author} {\bibfnamefont {C.}~\bibnamefont {Nayak}}, \ and\ \bibinfo
  {author} {\bibfnamefont {N.~Y.}\ \bibnamefont {Yao}},\ }\href {\doibase
  10.1146/annurev-conmatphys-031119-050658} {\bibfield  {journal} {\bibinfo
  {journal} {Annu. Rev. Condens. Matter Phys.}\ }\textbf {\bibinfo {volume}
  {11}},\ \bibinfo {pages} {467} (\bibinfo {year} {2020})}\BibitemShut
  {NoStop}%
\bibitem [{\citenamefont {Zaletel}\ \emph {et~al.}(2023)\citenamefont
  {Zaletel}, \citenamefont {Lukin}, \citenamefont {Monroe}, \citenamefont
  {Nayak}, \citenamefont {Wilczek},\ and\ \citenamefont {Yao}}]{Zaletel2023}%
  \BibitemOpen
  \bibfield  {author} {\bibinfo {author} {\bibfnamefont {M.~P.}\ \bibnamefont
  {Zaletel}}, \bibinfo {author} {\bibfnamefont {M.}~\bibnamefont {Lukin}},
  \bibinfo {author} {\bibfnamefont {C.}~\bibnamefont {Monroe}}, \bibinfo
  {author} {\bibfnamefont {C.}~\bibnamefont {Nayak}}, \bibinfo {author}
  {\bibfnamefont {F.}~\bibnamefont {Wilczek}}, \ and\ \bibinfo {author}
  {\bibfnamefont {N.~Y.}\ \bibnamefont {Yao}},\ }\href {\doibase
  10.1103/RevModPhys.95.031001} {\bibfield  {journal} {\bibinfo  {journal}
  {Rev. Mod. Phys.}\ }\textbf {\bibinfo {volume} {95}},\ \bibinfo {pages}
  {031001} (\bibinfo {year} {2023})}\BibitemShut {NoStop}%
\bibitem [{\citenamefont {Khemani}\ \emph {et~al.}(2016)\citenamefont
  {Khemani}, \citenamefont {Lazarides}, \citenamefont {Moessner},\ and\
  \citenamefont {Sondhi}}]{Khemani2016}%
  \BibitemOpen
  \bibfield  {author} {\bibinfo {author} {\bibfnamefont {V.}~\bibnamefont
  {Khemani}}, \bibinfo {author} {\bibfnamefont {A.}~\bibnamefont {Lazarides}},
  \bibinfo {author} {\bibfnamefont {R.}~\bibnamefont {Moessner}}, \ and\
  \bibinfo {author} {\bibfnamefont {S.~L.}\ \bibnamefont {Sondhi}},\ }\href
  {\doibase 10.1103/PhysRevLett.116.250401} {\bibfield  {journal} {\bibinfo
  {journal} {Phys. Rev. Lett.}\ }\textbf {\bibinfo {volume} {116}},\ \bibinfo
  {pages} {250401} (\bibinfo {year} {2016})}\BibitemShut {NoStop}%
\bibitem [{\citenamefont {Else}\ \emph {et~al.}(2016)\citenamefont {Else},
  \citenamefont {Bauer},\ and\ \citenamefont {Nayak}}]{Else2016}%
  \BibitemOpen
  \bibfield  {author} {\bibinfo {author} {\bibfnamefont {D.~V.}\ \bibnamefont
  {Else}}, \bibinfo {author} {\bibfnamefont {B.}~\bibnamefont {Bauer}}, \ and\
  \bibinfo {author} {\bibfnamefont {C.}~\bibnamefont {Nayak}},\ }\href
  {\doibase 10.1103/PhysRevLett.117.090402} {\bibfield  {journal} {\bibinfo
  {journal} {Phys. Rev. Lett.}\ }\textbf {\bibinfo {volume} {117}},\ \bibinfo
  {pages} {090402} (\bibinfo {year} {2016})}\BibitemShut {NoStop}%
\bibitem [{\citenamefont {von Keyserlingk}\ \emph {et~al.}(2016)\citenamefont
  {von Keyserlingk}, \citenamefont {Khemani},\ and\ \citenamefont
  {Sondhi}}]{Keyserlingk2016a}%
  \BibitemOpen
  \bibfield  {author} {\bibinfo {author} {\bibfnamefont {C.~W.}\ \bibnamefont
  {von Keyserlingk}}, \bibinfo {author} {\bibfnamefont {V.}~\bibnamefont
  {Khemani}}, \ and\ \bibinfo {author} {\bibfnamefont {S.~L.}\ \bibnamefont
  {Sondhi}},\ }\href {\doibase 10.1103/PhysRevB.94.085112} {\bibfield
  {journal} {\bibinfo  {journal} {Phys. Rev. B}\ }\textbf {\bibinfo {volume}
  {94}},\ \bibinfo {pages} {085112} (\bibinfo {year} {2016})}\BibitemShut
  {NoStop}%
\bibitem [{\citenamefont {Surace}\ \emph {et~al.}(2019)\citenamefont {Surace},
  \citenamefont {Russomanno}, \citenamefont {Dalmonte}, \citenamefont {Silva},
  \citenamefont {Fazio},\ and\ \citenamefont {Iemini}}]{Surace2019}%
  \BibitemOpen
  \bibfield  {author} {\bibinfo {author} {\bibfnamefont {F.~M.}\ \bibnamefont
  {Surace}}, \bibinfo {author} {\bibfnamefont {A.}~\bibnamefont {Russomanno}},
  \bibinfo {author} {\bibfnamefont {M.}~\bibnamefont {Dalmonte}}, \bibinfo
  {author} {\bibfnamefont {A.}~\bibnamefont {Silva}}, \bibinfo {author}
  {\bibfnamefont {R.}~\bibnamefont {Fazio}}, \ and\ \bibinfo {author}
  {\bibfnamefont {F.}~\bibnamefont {Iemini}},\ }\href {\doibase
  10.1103/PhysRevB.99.104303} {\bibfield  {journal} {\bibinfo  {journal} {Phys.
  Rev. B}\ }\textbf {\bibinfo {volume} {99}},\ \bibinfo {pages} {104303}
  (\bibinfo {year} {2019})}\BibitemShut {NoStop}%
\bibitem [{\citenamefont {Lazarides}\ \emph {et~al.}(2015)\citenamefont
  {Lazarides}, \citenamefont {Das},\ and\ \citenamefont
  {Moessner}}]{Lazarides2015}%
  \BibitemOpen
  \bibfield  {author} {\bibinfo {author} {\bibfnamefont {A.}~\bibnamefont
  {Lazarides}}, \bibinfo {author} {\bibfnamefont {A.}~\bibnamefont {Das}}, \
  and\ \bibinfo {author} {\bibfnamefont {R.}~\bibnamefont {Moessner}},\ }\href
  {\doibase 10.1103/PhysRevLett.115.030402} {\bibfield  {journal} {\bibinfo
  {journal} {Phys. Rev. Lett.}\ }\textbf {\bibinfo {volume} {115}},\ \bibinfo
  {pages} {030402} (\bibinfo {year} {2015})}\BibitemShut {NoStop}%
\bibitem [{\citenamefont {Yao}\ \emph {et~al.}(2017)\citenamefont {Yao},
  \citenamefont {Potter}, \citenamefont {Potirniche},\ and\ \citenamefont
  {Vishwanath}}]{Yao2017}%
  \BibitemOpen
  \bibfield  {author} {\bibinfo {author} {\bibfnamefont {N.~Y.}\ \bibnamefont
  {Yao}}, \bibinfo {author} {\bibfnamefont {A.~C.}\ \bibnamefont {Potter}},
  \bibinfo {author} {\bibfnamefont {I.-D.}\ \bibnamefont {Potirniche}}, \ and\
  \bibinfo {author} {\bibfnamefont {A.}~\bibnamefont {Vishwanath}},\ }\href
  {\doibase 10.1103/PhysRevLett.118.030401} {\bibfield  {journal} {\bibinfo
  {journal} {Phys. Rev. Lett.}\ }\textbf {\bibinfo {volume} {118}},\ \bibinfo
  {pages} {030401} (\bibinfo {year} {2017})}\BibitemShut {NoStop}%
\bibitem [{\citenamefont {Ponte}\ \emph {et~al.}(2015)\citenamefont {Ponte},
  \citenamefont {Papi\ifmmode~\acute{c}\else \'{c}\fi{}}, \citenamefont
  {Huveneers},\ and\ \citenamefont {Abanin}}]{Ponte2015}%
  \BibitemOpen
  \bibfield  {author} {\bibinfo {author} {\bibfnamefont {P.}~\bibnamefont
  {Ponte}}, \bibinfo {author} {\bibfnamefont {Z.}~\bibnamefont
  {Papi\ifmmode~\acute{c}\else \'{c}\fi{}}}, \bibinfo {author} {\bibfnamefont
  {F.~m.~c.}\ \bibnamefont {Huveneers}}, \ and\ \bibinfo {author}
  {\bibfnamefont {D.~A.}\ \bibnamefont {Abanin}},\ }\href {\doibase
  10.1103/PhysRevLett.114.140401} {\bibfield  {journal} {\bibinfo  {journal}
  {Phys. Rev. Lett.}\ }\textbf {\bibinfo {volume} {114}},\ \bibinfo {pages}
  {140401} (\bibinfo {year} {2015})}\BibitemShut {NoStop}%
\bibitem [{\citenamefont {Bairey}\ \emph {et~al.}(2017)\citenamefont {Bairey},
  \citenamefont {Refael},\ and\ \citenamefont {Lindner}}]{Bairey2017}%
  \BibitemOpen
  \bibfield  {author} {\bibinfo {author} {\bibfnamefont {E.}~\bibnamefont
  {Bairey}}, \bibinfo {author} {\bibfnamefont {G.}~\bibnamefont {Refael}}, \
  and\ \bibinfo {author} {\bibfnamefont {N.~H.}\ \bibnamefont {Lindner}},\
  }\href {\doibase 10.1103/PhysRevB.96.020201} {\bibfield  {journal} {\bibinfo
  {journal} {Phys. Rev. B}\ }\textbf {\bibinfo {volume} {96}},\ \bibinfo
  {pages} {020201} (\bibinfo {year} {2017})}\BibitemShut {NoStop}%
\bibitem [{\citenamefont {Sonner}\ \emph {et~al.}(2021)\citenamefont {Sonner},
  \citenamefont {Serbyn}, \citenamefont {Papi\ifmmode~\acute{c}\else
  \'{c}\fi{}},\ and\ \citenamefont {Abanin}}]{Sonner2021}%
  \BibitemOpen
  \bibfield  {author} {\bibinfo {author} {\bibfnamefont {M.}~\bibnamefont
  {Sonner}}, \bibinfo {author} {\bibfnamefont {M.}~\bibnamefont {Serbyn}},
  \bibinfo {author} {\bibfnamefont {Z.}~\bibnamefont
  {Papi\ifmmode~\acute{c}\else \'{c}\fi{}}}, \ and\ \bibinfo {author}
  {\bibfnamefont {D.~A.}\ \bibnamefont {Abanin}},\ }\href {\doibase
  10.1103/PhysRevB.104.L081112} {\bibfield  {journal} {\bibinfo  {journal}
  {Phys. Rev. B}\ }\textbf {\bibinfo {volume} {104}},\ \bibinfo {pages}
  {L081112} (\bibinfo {year} {2021})}\BibitemShut {NoStop}%
\bibitem [{\citenamefont {Sierant}\ \emph {et~al.}(2023)\citenamefont
  {Sierant}, \citenamefont {Lewenstein}, \citenamefont {Scardicchio},\ and\
  \citenamefont {Zakrzewski}}]{Sierant2023}%
  \BibitemOpen
  \bibfield  {author} {\bibinfo {author} {\bibfnamefont {P.}~\bibnamefont
  {Sierant}}, \bibinfo {author} {\bibfnamefont {M.}~\bibnamefont {Lewenstein}},
  \bibinfo {author} {\bibfnamefont {A.}~\bibnamefont {Scardicchio}}, \ and\
  \bibinfo {author} {\bibfnamefont {J.}~\bibnamefont {Zakrzewski}},\ }\href
  {\doibase 10.1103/PhysRevB.107.115132} {\bibfield  {journal} {\bibinfo
  {journal} {Phys. Rev. B}\ }\textbf {\bibinfo {volume} {107}},\ \bibinfo
  {pages} {115132} (\bibinfo {year} {2023})}\BibitemShut {NoStop}%
\bibitem [{\citenamefont {Lieb}\ \emph {et~al.}(1961)\citenamefont {Lieb},
  \citenamefont {Schultz},\ and\ \citenamefont {Mattis}}]{Lieb1961}%
  \BibitemOpen
  \bibfield  {author} {\bibinfo {author} {\bibfnamefont {E.}~\bibnamefont
  {Lieb}}, \bibinfo {author} {\bibfnamefont {T.}~\bibnamefont {Schultz}}, \
  and\ \bibinfo {author} {\bibfnamefont {D.}~\bibnamefont {Mattis}},\ }\href
  {\doibase https://doi.org/10.1016/0003-4916(61)90115-4} {\bibfield  {journal}
  {\bibinfo  {journal} {Ann. Phys}\ }\textbf {\bibinfo {volume} {16}},\
  \bibinfo {pages} {407} (\bibinfo {year} {1961})}\BibitemShut {NoStop}%
\bibitem [{\citenamefont {Pfeuty}(1979)}]{Pfeuty1979}%
  \BibitemOpen
  \bibfield  {author} {\bibinfo {author} {\bibfnamefont {P.}~\bibnamefont
  {Pfeuty}},\ }\href {\doibase 10.1016/0375-9601(79)90017-3} {\bibfield
  {journal} {\bibinfo  {journal} {Phys. Lett. A}\ }\textbf {\bibinfo {volume}
  {72}},\ \bibinfo {pages} {245} (\bibinfo {year} {1979})}\BibitemShut
  {NoStop}%
\bibitem [{\citenamefont {Imry}(1986)}]{Imry1986}%
  \BibitemOpen
  \bibfield  {author} {\bibinfo {author} {\bibfnamefont {Y.}~\bibnamefont
  {Imry}},\ }\href {\doibase 10.1209/0295-5075/1/5/008} {\bibfield  {journal}
  {\bibinfo  {journal} {Europhys. Lett.}\ }\textbf {\bibinfo {volume} {1}},\
  \bibinfo {pages} {249} (\bibinfo {year} {1986})}\BibitemShut {NoStop}%
\bibitem [{\citenamefont {Titov}\ \emph {et~al.}(1997)\citenamefont {Titov},
  \citenamefont {Braun},\ and\ \citenamefont {Fyodorov}}]{Titov1997}%
  \BibitemOpen
  \bibfield  {author} {\bibinfo {author} {\bibfnamefont {M.}~\bibnamefont
  {Titov}}, \bibinfo {author} {\bibfnamefont {D.}~\bibnamefont {Braun}}, \ and\
  \bibinfo {author} {\bibfnamefont {Y.~V.}\ \bibnamefont {Fyodorov}},\ }\href
  {\doibase 10.1088/0305-4470/30/10/007} {\bibfield  {journal} {\bibinfo
  {journal} {J. Phys. A: Math. and Gen.}\ }\textbf {\bibinfo {volume} {30}},\
  \bibinfo {pages} {L339} (\bibinfo {year} {1997})}\BibitemShut {NoStop}%
\bibitem [{\citenamefont {Brouwer}\ \emph {et~al.}(2011)\citenamefont
  {Brouwer}, \citenamefont {Duckheim}, \citenamefont {Romito},\ and\
  \citenamefont {von Oppen}}]{Brouwer2011}%
  \BibitemOpen
  \bibfield  {author} {\bibinfo {author} {\bibfnamefont {P.~W.}\ \bibnamefont
  {Brouwer}}, \bibinfo {author} {\bibfnamefont {M.}~\bibnamefont {Duckheim}},
  \bibinfo {author} {\bibfnamefont {A.}~\bibnamefont {Romito}}, \ and\ \bibinfo
  {author} {\bibfnamefont {F.}~\bibnamefont {von Oppen}},\ }\href {\doibase
  10.1103/PhysRevLett.107.196804} {\bibfield  {journal} {\bibinfo  {journal}
  {Phys. Rev. Lett.}\ }\textbf {\bibinfo {volume} {107}},\ \bibinfo {pages}
  {196804} (\bibinfo {year} {2011})}\BibitemShut {NoStop}%
\bibitem [{\citenamefont {Asmussen}\ \emph {et~al.}(2016)\citenamefont
  {Asmussen}, \citenamefont {Jensen},\ and\ \citenamefont
  {Rojas-Nandayapa}}]{Asmussen2016}%
  \BibitemOpen
  \bibfield  {author} {\bibinfo {author} {\bibfnamefont {S.}~\bibnamefont
  {Asmussen}}, \bibinfo {author} {\bibfnamefont {J.~L.}\ \bibnamefont
  {Jensen}}, \ and\ \bibinfo {author} {\bibfnamefont {L.}~\bibnamefont
  {Rojas-Nandayapa}},\ }\href {\doibase 10.1007/s11009-014-9430-7} {\bibfield
  {journal} {\bibinfo  {journal} {Methodol. Comput. Appl. Probab.}\ }\textbf
  {\bibinfo {volume} {18}},\ \bibinfo {pages} {441} (\bibinfo {year}
  {2016})}\BibitemShut {NoStop}%
\bibitem [{\citenamefont {Schmid}\ \emph {et~al.}(2024)\citenamefont {Schmid},
  \citenamefont {Penner}, \citenamefont {Yang}, \citenamefont {Glazman},\ and\
  \citenamefont {von Oppen}}]{schmid2024robust}%
  \BibitemOpen
  \bibfield  {author} {\bibinfo {author} {\bibfnamefont {H.}~\bibnamefont
  {Schmid}}, \bibinfo {author} {\bibfnamefont {A.-G.}\ \bibnamefont {Penner}},
  \bibinfo {author} {\bibfnamefont {K.}~\bibnamefont {Yang}}, \bibinfo {author}
  {\bibfnamefont {L.}~\bibnamefont {Glazman}}, \ and\ \bibinfo {author}
  {\bibfnamefont {F.}~\bibnamefont {von Oppen}},\ }\href {\doibase
  10.1103/PhysRevLett.132.210401} {\bibfield  {journal} {\bibinfo  {journal}
  {Phys. Rev. Lett.}\ }\textbf {\bibinfo {volume} {132}},\ \bibinfo {pages}
  {210401} (\bibinfo {year} {2024})}\BibitemShut {NoStop}%
\bibitem [{\citenamefont {Thouless}(1977)}]{Thouless1977}%
  \BibitemOpen
  \bibfield  {author} {\bibinfo {author} {\bibfnamefont {D.~J.}\ \bibnamefont
  {Thouless}},\ }\href {\doibase 10.1103/PhysRevLett.39.1167} {\bibfield
  {journal} {\bibinfo  {journal} {Phys. Rev. Lett.}\ }\textbf {\bibinfo
  {volume} {39}},\ \bibinfo {pages} {1167} (\bibinfo {year}
  {1977})}\BibitemShut {NoStop}%
\bibitem [{\citenamefont {Oseledec}(1968)}]{Oseledec1968}%
  \BibitemOpen
  \bibfield  {author} {\bibinfo {author} {\bibfnamefont {V.~I.}\ \bibnamefont
  {Oseledec}},\ }\href {https://cir.nii.ac.jp/crid/1571135650268691328}
  {\bibfield  {journal} {\bibinfo  {journal} {Trans. Moscow Math. Soc.}\
  }\textbf {\bibinfo {volume} {19}},\ \bibinfo {pages} {197} (\bibinfo {year}
  {1968})}\BibitemShut {NoStop}%
\bibitem [{\citenamefont {Thouless}(1972)}]{Thouless1972}%
  \BibitemOpen
  \bibfield  {author} {\bibinfo {author} {\bibfnamefont {D.~J.}\ \bibnamefont
  {Thouless}},\ }\href {\doibase 10.1088/0022-3719/5/1/010} {\bibfield
  {journal} {\bibinfo  {journal} {J. Phys. C: Sol. St. Phys.}\ }\textbf
  {\bibinfo {volume} {5}},\ \bibinfo {pages} {77} (\bibinfo {year}
  {1972})}\BibitemShut {NoStop}%
\bibitem [{\citenamefont {Thouless}(1979)}]{Thouless1979}%
  \BibitemOpen
  \bibfield  {author} {\bibinfo {author} {\bibfnamefont {D.~J.}\ \bibnamefont
  {Thouless}},\ }\enquote {\bibinfo {title} {Percolation and localization},}\
  in\ \href {https://books.google.de/books?id=RTNqDQAAQBAJ} {\emph {\bibinfo
  {booktitle} {Ill-condensed {Matter}}}},\ \bibinfo {series and number} {Ecole
  d'ete de physique theoretique Les Houches},\ \bibinfo {editor} {edited by\
  \bibinfo {editor} {\bibfnamefont {R.}~\bibnamefont {Balian}}, \bibinfo
  {editor} {\bibfnamefont {R.}~\bibnamefont {Maynard}}, \ and\ \bibinfo
  {editor} {\bibfnamefont {G.}~\bibnamefont {Toulouse}}}\ (\bibinfo
  {publisher} {North-Holland Publishing Company},\ \bibinfo {year}
  {1979})\BibitemShut {NoStop}%
\bibitem [{\citenamefont {Sahu}\ and\ \citenamefont {Iemini}(2024)}]{Sahu2024}%
  \BibitemOpen
  \bibfield  {author} {\bibinfo {author} {\bibfnamefont {H.}~\bibnamefont
  {Sahu}}\ and\ \bibinfo {author} {\bibfnamefont {F.}~\bibnamefont {Iemini}},\
  }\href {https://arxiv.org/abs/2411.13469} {\enquote {\bibinfo {title}
  {Information scrambling and entanglement dynamics in floquet time
  crystals},}\ } (\bibinfo {year} {2024}),\ \Eprint
  {http://arxiv.org/abs/2411.13469} {arXiv:2411.13469} \BibitemShut {NoStop}%
\bibitem [{\citenamefont {Pizzi}\ \emph {et~al.}(2019)\citenamefont {Pizzi},
  \citenamefont {Knolle},\ and\ \citenamefont {Nunnenkamp}}]{Pizzi2019}%
  \BibitemOpen
  \bibfield  {author} {\bibinfo {author} {\bibfnamefont {A.}~\bibnamefont
  {Pizzi}}, \bibinfo {author} {\bibfnamefont {J.}~\bibnamefont {Knolle}}, \
  and\ \bibinfo {author} {\bibfnamefont {A.}~\bibnamefont {Nunnenkamp}},\
  }\href {\doibase 10.1103/PhysRevLett.123.150601} {\bibfield  {journal}
  {\bibinfo  {journal} {Phys. Rev. Lett.}\ }\textbf {\bibinfo {volume} {123}},\
  \bibinfo {pages} {150601} (\bibinfo {year} {2019})}\BibitemShut {NoStop}%
\bibitem [{\citenamefont {Pizzi}\ \emph {et~al.}(2021)\citenamefont {Pizzi},
  \citenamefont {Knolle},\ and\ \citenamefont {Nunnenkamp}}]{Pizzi2021}%
  \BibitemOpen
  \bibfield  {author} {\bibinfo {author} {\bibfnamefont {A.}~\bibnamefont
  {Pizzi}}, \bibinfo {author} {\bibfnamefont {J.}~\bibnamefont {Knolle}}, \
  and\ \bibinfo {author} {\bibfnamefont {A.}~\bibnamefont {Nunnenkamp}},\
  }\href {https://doi.org/10.1038/s41467-021-22583-5} {\bibfield  {journal}
  {\bibinfo  {journal} {Nat Comm}\ }\textbf {\bibinfo {volume} {12}},\ \bibinfo
  {pages} {2341} (\bibinfo {year} {2021})}\BibitemShut {NoStop}%
\bibitem [{\citenamefont {Giachetti}\ \emph {et~al.}(2023)\citenamefont
  {Giachetti}, \citenamefont {Solfanelli}, \citenamefont {Correale},\ and\
  \citenamefont {Defenu}}]{Giachetti2023}%
  \BibitemOpen
  \bibfield  {author} {\bibinfo {author} {\bibfnamefont {G.}~\bibnamefont
  {Giachetti}}, \bibinfo {author} {\bibfnamefont {A.}~\bibnamefont
  {Solfanelli}}, \bibinfo {author} {\bibfnamefont {L.}~\bibnamefont
  {Correale}}, \ and\ \bibinfo {author} {\bibfnamefont {N.}~\bibnamefont
  {Defenu}},\ }\href {\doibase 10.1103/PhysRevB.108.L140102} {\bibfield
  {journal} {\bibinfo  {journal} {Phys. Rev. B}\ }\textbf {\bibinfo {volume}
  {108}},\ \bibinfo {pages} {L140102} (\bibinfo {year} {2023})}\BibitemShut
  {NoStop}%
\bibitem [{\citenamefont {Iemini}\ \emph {et~al.}(2018)\citenamefont {Iemini},
  \citenamefont {Russomanno}, \citenamefont {Keeling}, \citenamefont
  {Schir\`o}, \citenamefont {Dalmonte},\ and\ \citenamefont
  {Fazio}}]{Iemini2018}%
  \BibitemOpen
  \bibfield  {author} {\bibinfo {author} {\bibfnamefont {F.}~\bibnamefont
  {Iemini}}, \bibinfo {author} {\bibfnamefont {A.}~\bibnamefont {Russomanno}},
  \bibinfo {author} {\bibfnamefont {J.}~\bibnamefont {Keeling}}, \bibinfo
  {author} {\bibfnamefont {M.}~\bibnamefont {Schir\`o}}, \bibinfo {author}
  {\bibfnamefont {M.}~\bibnamefont {Dalmonte}}, \ and\ \bibinfo {author}
  {\bibfnamefont {R.}~\bibnamefont {Fazio}},\ }\href {\doibase
  10.1103/PhysRevLett.121.035301} {\bibfield  {journal} {\bibinfo  {journal}
  {Phys. Rev. Lett.}\ }\textbf {\bibinfo {volume} {121}},\ \bibinfo {pages}
  {035301} (\bibinfo {year} {2018})}\BibitemShut {NoStop}%
\bibitem [{\citenamefont {Tucker}\ \emph {et~al.}(2018)\citenamefont {Tucker},
  \citenamefont {Zhu}, \citenamefont {Lewis-Swan}, \citenamefont {Marino},
  \citenamefont {Jimenez}, \citenamefont {Restrepo},\ and\ \citenamefont
  {Rey}}]{Tucker2018}%
  \BibitemOpen
  \bibfield  {author} {\bibinfo {author} {\bibfnamefont {K.}~\bibnamefont
  {Tucker}}, \bibinfo {author} {\bibfnamefont {B.}~\bibnamefont {Zhu}},
  \bibinfo {author} {\bibfnamefont {R.~J.}\ \bibnamefont {Lewis-Swan}},
  \bibinfo {author} {\bibfnamefont {J.}~\bibnamefont {Marino}}, \bibinfo
  {author} {\bibfnamefont {F.}~\bibnamefont {Jimenez}}, \bibinfo {author}
  {\bibfnamefont {J.~G.}\ \bibnamefont {Restrepo}}, \ and\ \bibinfo {author}
  {\bibfnamefont {A.~M.}\ \bibnamefont {Rey}},\ }\href {\doibase
  10.1088/1367-2630/aaf18b} {\bibfield  {journal} {\bibinfo  {journal} {New J.
  Phys.}\ }\textbf {\bibinfo {volume} {20}},\ \bibinfo {pages} {123003}
  (\bibinfo {year} {2018})}\BibitemShut {NoStop}%
\bibitem [{\citenamefont {Zhu}\ \emph {et~al.}(2019)\citenamefont {Zhu},
  \citenamefont {Marino}, \citenamefont {Yao}, \citenamefont {Lukin},\ and\
  \citenamefont {Demler}}]{Zhu2019}%
  \BibitemOpen
  \bibfield  {author} {\bibinfo {author} {\bibfnamefont {B.}~\bibnamefont
  {Zhu}}, \bibinfo {author} {\bibfnamefont {J.}~\bibnamefont {Marino}},
  \bibinfo {author} {\bibfnamefont {N.~Y.}\ \bibnamefont {Yao}}, \bibinfo
  {author} {\bibfnamefont {M.~D.}\ \bibnamefont {Lukin}}, \ and\ \bibinfo
  {author} {\bibfnamefont {E.~A.}\ \bibnamefont {Demler}},\ }\href
  {https://dx.doi.org/10.1088/1367-2630/ab2afe} {\bibfield  {journal} {\bibinfo
   {journal} {New J. Phys.}\ }\textbf {\bibinfo {volume} {21}},\ \bibinfo
  {pages} {073028} (\bibinfo {year} {2019})}\BibitemShut {NoStop}%
\bibitem [{\citenamefont {Gambetta}\ \emph {et~al.}(2019)\citenamefont
  {Gambetta}, \citenamefont {Carollo}, \citenamefont {Marcuzzi}, \citenamefont
  {Garrahan},\ and\ \citenamefont {Lesanovsky}}]{Gambetta2019}%
  \BibitemOpen
  \bibfield  {author} {\bibinfo {author} {\bibfnamefont {F.~M.}\ \bibnamefont
  {Gambetta}}, \bibinfo {author} {\bibfnamefont {F.}~\bibnamefont {Carollo}},
  \bibinfo {author} {\bibfnamefont {M.}~\bibnamefont {Marcuzzi}}, \bibinfo
  {author} {\bibfnamefont {J.~P.}\ \bibnamefont {Garrahan}}, \ and\ \bibinfo
  {author} {\bibfnamefont {I.}~\bibnamefont {Lesanovsky}},\ }\href {\doibase
  10.1103/PhysRevLett.122.015701} {\bibfield  {journal} {\bibinfo  {journal}
  {Phys. Rev. Lett.}\ }\textbf {\bibinfo {volume} {122}},\ \bibinfo {pages}
  {015701} (\bibinfo {year} {2019})}\BibitemShut {NoStop}%
\bibitem [{\citenamefont {Booker}\ \emph {et~al.}(2020)\citenamefont {Booker},
  \citenamefont {Buča},\ and\ \citenamefont {Jaksch}}]{Booker2020}%
  \BibitemOpen
  \bibfield  {author} {\bibinfo {author} {\bibfnamefont {C.}~\bibnamefont
  {Booker}}, \bibinfo {author} {\bibfnamefont {B.}~\bibnamefont {Buča}}, \
  and\ \bibinfo {author} {\bibfnamefont {D.}~\bibnamefont {Jaksch}},\ }\href
  {\doibase 10.1088/1367-2630/ababc4} {\bibfield  {journal} {\bibinfo
  {journal} {New J. Phys.}\ }\textbf {\bibinfo {volume} {22}},\ \bibinfo
  {pages} {085007} (\bibinfo {year} {2020})}\BibitemShut {NoStop}%
\bibitem [{\citenamefont {Krishna}\ \emph {et~al.}(2023)\citenamefont
  {Krishna}, \citenamefont {Solanki}, \citenamefont
  {Hajdu\ifmmode~\check{s}\else \v{s}\fi{}ek},\ and\ \citenamefont
  {Vinjanampathy}}]{Krishna2023}%
  \BibitemOpen
  \bibfield  {author} {\bibinfo {author} {\bibfnamefont {M.}~\bibnamefont
  {Krishna}}, \bibinfo {author} {\bibfnamefont {P.}~\bibnamefont {Solanki}},
  \bibinfo {author} {\bibfnamefont {M.}~\bibnamefont
  {Hajdu\ifmmode~\check{s}\else \v{s}\fi{}ek}}, \ and\ \bibinfo {author}
  {\bibfnamefont {S.}~\bibnamefont {Vinjanampathy}},\ }\href {\doibase
  10.1103/PhysRevLett.130.150401} {\bibfield  {journal} {\bibinfo  {journal}
  {Phys. Rev. Lett.}\ }\textbf {\bibinfo {volume} {130}},\ \bibinfo {pages}
  {150401} (\bibinfo {year} {2023})}\BibitemShut {NoStop}%
\bibitem [{\citenamefont {van Nieuwenburg}\ \emph {et~al.}(2019)\citenamefont
  {van Nieuwenburg}, \citenamefont {Baum},\ and\ \citenamefont
  {Refael}}]{Van2019}%
  \BibitemOpen
  \bibfield  {author} {\bibinfo {author} {\bibfnamefont {E.}~\bibnamefont {van
  Nieuwenburg}}, \bibinfo {author} {\bibfnamefont {Y.}~\bibnamefont {Baum}}, \
  and\ \bibinfo {author} {\bibfnamefont {G.}~\bibnamefont {Refael}},\ }\href
  {www.pnas.org/doi/full/10.1073/pnas.1819316116} {\bibfield  {journal}
  {\bibinfo  {journal} {Proc. Natl. Acad. Sci.}\ }\textbf {\bibinfo {volume}
  {116}},\ \bibinfo {pages} {9269} (\bibinfo {year} {2019})}\BibitemShut
  {NoStop}%
\bibitem [{\citenamefont {Kshetrimayum}\ \emph {et~al.}(2020)\citenamefont
  {Kshetrimayum}, \citenamefont {Eisert},\ and\ \citenamefont
  {Kennes}}]{Kshetrimayum2020}%
  \BibitemOpen
  \bibfield  {author} {\bibinfo {author} {\bibfnamefont {A.}~\bibnamefont
  {Kshetrimayum}}, \bibinfo {author} {\bibfnamefont {J.}~\bibnamefont
  {Eisert}}, \ and\ \bibinfo {author} {\bibfnamefont {D.~M.}\ \bibnamefont
  {Kennes}},\ }\href {\doibase 10.1103/PhysRevB.102.195116} {\bibfield
  {journal} {\bibinfo  {journal} {Phys. Rev. B}\ }\textbf {\bibinfo {volume}
  {102}},\ \bibinfo {pages} {195116} (\bibinfo {year} {2020})}\BibitemShut
  {NoStop}%
\bibitem [{\citenamefont {Liu}\ \emph {et~al.}(2023)\citenamefont {Liu},
  \citenamefont {Zhang}, \citenamefont {Hsieh}, \citenamefont {Zhang},\ and\
  \citenamefont {Yao}}]{Liu2023}%
  \BibitemOpen
  \bibfield  {author} {\bibinfo {author} {\bibfnamefont {S.}~\bibnamefont
  {Liu}}, \bibinfo {author} {\bibfnamefont {S.-X.}\ \bibnamefont {Zhang}},
  \bibinfo {author} {\bibfnamefont {C.-Y.}\ \bibnamefont {Hsieh}}, \bibinfo
  {author} {\bibfnamefont {S.}~\bibnamefont {Zhang}}, \ and\ \bibinfo {author}
  {\bibfnamefont {H.}~\bibnamefont {Yao}},\ }\href {\doibase
  10.1103/PhysRevLett.130.120403} {\bibfield  {journal} {\bibinfo  {journal}
  {Phys. Rev. Lett.}\ }\textbf {\bibinfo {volume} {130}},\ \bibinfo {pages}
  {120403} (\bibinfo {year} {2023})}\BibitemShut {NoStop}%
\bibitem [{\citenamefont {Bennett}\ \emph {et~al.}(2020)\citenamefont
  {Bennett}, \citenamefont {Melchers},\ and\ \citenamefont
  {Proppe}}]{Bennett2020}%
  \BibitemOpen
  \bibfield  {author} {\bibinfo {author} {\bibfnamefont {L.}~\bibnamefont
  {Bennett}}, \bibinfo {author} {\bibfnamefont {B.}~\bibnamefont {Melchers}}, \
  and\ \bibinfo {author} {\bibfnamefont {B.}~\bibnamefont {Proppe}},\ }\href
  {http://dx.doi.org/10.17169/refubium-26754} {\enquote {\bibinfo {title}
  {{Curta: A General-purpose High-Performance Computer at ZEDAT, Freie
  Universit{\"a}t Berlin}},}\ } (\bibinfo {year} {2020})\BibitemShut {NoStop}%
\end{thebibliography}

%

\end{document}